\documentclass[preprint,showpacs,showkeys]{revtex4}

\usepackage{amssymb}

\usepackage{graphicx}
\usepackage{dcolumn}
\usepackage{bm}

\begin{document}

\title{Scaling, Self-similarity and Superposition}
\author{Kaz{\i}m Yavuz Ek\c{s}i}
\email{eksi@itu.edu.tr}
\affiliation{Istanbul Technical University, Faculty of Science and Letters, \\
Dept. of Physics Engineering, Maslak 34469, Istanbul, TURKEY }
\date{\today }

\begin{abstract}

A novel procedure for the nonlinear superposition of two self-similar 
solutions of the heat conduction equation with power-law nonlinearity 
is introduced. It is shown how the boundary conditions of the 
superposed state conflicts with self-similarity, rendering the
nonlinearly superposed state to be a non-exact solution. 
It is argued that the nonlinearity couples with the presence of the scale so 
that the superposition in the linear case can give an exact solution. 

\end{abstract}



\maketitle

\section{Introduction}

Phenomena exhibiting self-similarity at different time and/or spatial scales 
are ubiquitous in nature \cite{bar96}.
Self-similarity techniques exploit such symmetries 
for reducing the number of variables for describing the system. This also
is the underlying idea in dimensional analysis \cite{ray15a,buck15}.  
Such techniques had much been explored in 
fluid phenomena \cite{bar72} and, more recently, in optics \cite{dudley}.
Self-similar solutions can only be constructed in the absence of scales having 
the dimensions of independent variables. As a result they are always endowed with extreme boundary
conditions (BCs) representing
the intermediate asymptotic behavior \cite{bar96} of a system away from the initial 
conditions and boundaries.

A PDE may have more than one
self-similar solution each corresponding to a specific BC.
If the underlying equation is linear, its self-similar solutions
can be superposed to obtain a solution satisfying a realistic boundary condition. 
Superposition of the solutions in the nonlinear case is not 
possible, but as is shown here, there is still an approximate symmetry to be exploited, 
leading to a procedure 
for nonlinear superposition of self-similar solutions.

\section{Heat Equation and Self-similar solutions}

As an example for illustrating the procedure, consider the nonlinear
diffusion equation 
\begin{equation}
\frac{\partial \theta }{\partial \tau }=\frac{\partial ^{2}\theta ^{n+1}}{%
\partial \xi ^{2}}  \label{heat1}
\end{equation}
which is the dimensionless form of the heat conduction equation with power-law nonlinearity \cite{zel67}.
This equation is also known as the porous medium equation \cite{vasq} describing the flow of an
isentropic gas through a porous medium. The equation is nonlinear for $n \neq 0$ and we call $n$ the nonlinearity parameter. 
This equation has two well-known self-similar solutions \cite{bar52,bar57,zel59}. 
The first solution, 
\begin{equation}
\theta(\xi,\tau) =\tau ^{-\frac{1}{n+1}}\left[ \xi \tau ^{-\frac{1}{2\left( n+1\right) 
}}\right] ^{\frac{1}{n+1}}\left[ 1-k_n\left( \xi \tau ^{-\frac{1}{2\left(
n+1\right) }}\right) ^{\frac{n+2}{n+1}}\right] ^{1/n},  \label{sol1}
\end{equation}
satisfies the Dirichlet BC $\theta(0,\tau)=0$, and the second solution,
\begin{equation}
\theta(\xi,\tau) =\tau ^{-\frac{1}{n+2}}\left[ 1-k_n\left( \xi \tau ^{-\frac{1}{n+2}%
}\right) ^{2}\right] ^{1/n},  \label{sol2}
\end{equation}
satisfies the Neumann BC $\theta'(0,\tau)=0$ where $\theta' \equiv d\theta/d\xi$. In both these solutions
\begin{equation}
k_n=\frac{n}{2( n+1) (n+2)}.  \label{k}
\end{equation}

The equation (\ref{heat1}) can equivalently be replaced by
two equations in conservative form. The first one is
\begin{equation}
\frac{\partial \theta }{\partial \tau }+\frac{\partial \Phi }{\partial \xi }=0
\label{con1}
\end{equation}%
where
\begin{equation}
\Phi(\xi,\tau) =-\frac{\partial \theta ^{n+1}}{\partial \xi }
\label{fick}
\end{equation}%
is the Fick's law in dimensionless form, and the second one is
\begin{equation}
\frac{\partial (\theta \xi )}{\partial \tau }+\frac{\partial \Gamma }{\partial \xi }=0
\label{con2}
\end{equation}%
where
\begin{equation}
\Gamma(\xi,\tau) =\Phi(\xi,\tau) \xi +\theta ^{n+1}
\label{stdy}
\end{equation}
It is clear from the Eqns.(\ref{con1})  and (\ref{con2}) that, in the steady state,
$\Phi $ and $\Gamma $ are the integration constants and solving equation (\ref{stdy})
for $\theta$ gives 
\begin{equation}
\theta(\xi) =(\Gamma -\Phi \xi ) ^{\frac{1}{n+1}},
\label{steady}
\end{equation}
which is the general solution in terms of $\Phi $ and $\Gamma$.

How do the time-dependent solutions given in Eqns. (\ref{sol1})  and (\ref{sol2})
look when written in terms of
$\Gamma(\tau) \equiv -\Gamma(0,\tau)$  and $\Phi(\tau) \equiv -\Phi(0,\tau)$ 
corresponding to the time dependent case? Here the negative sign is to make the 
flux leaving the system from the left boundary positive.
Using the divergence theorem for equation (\ref{con1}), it is possible to see from 
the first solution given in equation (\ref{sol1}) that
\begin{equation}
\Phi (\tau )=\frac{d}{d\tau}\int_0^{\infty} \theta d\xi  \label{flux}
\end{equation}%
declines as a power-law $\Phi (\tau )=-\tau ^{-\alpha}$ where
\begin{equation}
\alpha=1+\frac{1}{2(n+1)}  
\label{alfa}
\end{equation}
while 
\begin{equation}
\Gamma (\tau)=\frac{d}{d\tau}\int_0^{\infty} \xi \theta d\xi  \label{gama}
\end{equation}
vanishes ($\Gamma (\tau)=0$).

Similarly, it is possible to see from the second solution, 
given in equation (\ref{sol2}), that $\Gamma (\tau) = \tau^{-\beta}$ where
\begin{equation}
\beta = 1-\frac{1}{n+2} 
\label{beta}
\end{equation}
while $\Phi (\tau )=0$.

\section{The Solutions in Terms of Fluxes}

We write the first solution given in equation (\ref{sol1}) in terms of $\Phi (\tau)$ as
\begin{equation}
\theta =\left[ -\xi \Phi (\tau )\right] ^{\frac{1}{n+1}}\left\{ 1-k_n\frac{\xi
^2}{\tau }\left[ -\xi \Phi (\tau )\right] ^{\frac{1}{n+1}-1}\right\} ^{1/n}.
\label{sol1a}
\end{equation}
It is possible to show that this satisfies equation (\ref{heat1}) 
even if $\Phi(\tau )$ is multiplied with a constant $\Phi_0$ so that 
\begin{equation}
\Phi (\tau )=-\Phi _{0}\tau ^{-\alpha}.  \label{phi1a}
\end{equation}
The second solution given in 
equation (\ref{sol2}), can be written in terms of $\Gamma (\tau)$ as
\begin{equation}
\theta =\left[ \Gamma (\tau )\right] ^{\frac{1}{n+1}}\left\{ 1-k_n\frac{\xi
^{2}}{\tau}\left[ \Gamma (\tau )\right] ^{\frac{1}{n+1}-1}\right\} ^{1/n}
\label{sol2a}
\end{equation}
and it is possible to show that this satisfies equation (\ref{heat1}) even if 
$\Gamma(\tau)$ is defined as 
\begin{equation}
\Gamma (\tau )=\Gamma_0 \tau ^{-\beta}  \label{gamma1a}
\end{equation}
where $\Gamma_0$ is a constant. Note that there is a ``duality'' $\Gamma \leftrightarrow  -\xi \Phi$ between the 
two solutions (\ref{sol1a}) and (\ref{sol2a}).

The solutions for the linear case ($n=0$), using $\lim_{n\rightarrow 0}( 1+A n) ^{1/n}=e^{A}$, 
can be written, in terms of  $\Gamma$ and $\Phi$, as
\begin{equation}
\theta ( \xi ,\tau ) =-\xi \Phi (\tau ) e^{-\xi ^{2}/4\tau }
\label{sol3_lin}
\end{equation}
where $\Phi = -\Phi_0 \tau^{-3/2}$ and
\begin{equation}
\theta ( \xi ,\tau ) =\Gamma (\tau ) e^{-\xi ^{2}/4\tau }
\label{sol4_lin}
\end{equation}
where $\Gamma =\Gamma_0 \tau ^{-1/2}$, respectively. As the diffusion equation (\ref{heat1}) is linear for $n=0$,
these two can be superposed to give
\begin{equation}
\theta ( \xi ,\tau ) =[\Gamma (\tau )-\xi \Phi (\tau )] e^{-\xi ^{2}/4\tau },
\label{sol_lin}
\end{equation}
and this satisfies the boundary condition
$\theta(0,\tau) =\Gamma(\tau)$. Note that the steady state solution 
given in Eqn.(\ref{steady}) for the linear case ($n=0$) becomes
$\theta ( \xi ) =\Gamma -\xi \Phi$.
The term in the square brackets in Eqn.(\ref{sol_lin}) carries a form similar
to the steady state solution but $\Gamma$ and $\Phi$ are time-dependent in the former case.

In the linear case multiplying a solution with any constant can be absorbed into the constants $\Phi_0$
or $\Gamma_0$. In the nonlinear case multiplying the solution with a constant does not give a solution, 
but it is possible to gauge $\Phi_0$ or $\Gamma_0$ to obtain the similar effect. 
In other words, multiplying the flux with a constant in the general case reduces to multiplying the solution with a constant.

\section{Nonlinear Superposition}

The self-similar solutions given by equations (\ref{sol1a}) and (\ref{sol2a}) correspond to
extreme boundary conditons: According to the former, 
there is a sink at $\xi=0$, such that
all the heat carried to this boundary is totally absorbed. The latter solution 
describes the case in which
there is a perfect insulator at $\xi=0$ such that heat cannot leave the
system from this boundary. In the steady state, the solution of which is 
given in  equation(\ref{steady}), any
ratio between the integration constants $\Gamma$ and $\Phi$ is possible allowing for intermediate BCs. 
This is not the case with the self-similar solutions which can only be constructed in the absence of a scale: 
If $\theta$ and $\theta'$ are both finite at the boundary, then $l \sim \theta / \theta'$ is a length scale 
associated with the system. It is the aim of this paper to find the time dependent version of the general 
steady state solution given in equation (\ref{steady}) even though it may not be an exact solution.

The self-similar solutions given in equations (\ref{sol1a}) and (\ref{sol2a})
appear to be the time-dependent versions of the steady state solution 
given in equation (\ref{steady}) with $\Gamma =0$
and $\Phi =0$, respectively. We have written the solutions
given by equations (\ref{sol1a}) and (\ref{sol2a}) in a ``dual'' form suggesting a ``nonlinear superposition''
in which we add up not the solutions themselves but $\Gamma(\tau)$ and $-\xi \Phi(\tau)$ 
terms in separate corresponding parts of the solutions. This procedure gives
\begin{equation}
\theta(\xi,\tau) =[\Gamma(\tau) -\xi \Phi(\tau)]^{\frac{1}{n+1}}\left( 1-k_n\frac{\xi ^{2}}{\tau }%
[\Gamma(\tau) -\xi \Phi(\tau) ]^{\frac{1}{n+1}-1}\right) ^{1/n}  \label{unified}
\end{equation}
which reduces to equations (\ref{sol1a}) and (\ref{sol2a}) for $\Gamma_0=0$ and 
$\Phi_0=0$, respectively. 
This expression is endowed with the BC $\theta(0,\tau)=[\Gamma(\tau)]^{1/(n+1)}$
which can describe cases in which a fraction of the heat is absorbed at the boundary.

\begin{figure}
\centering
\includegraphics[width=75mm]{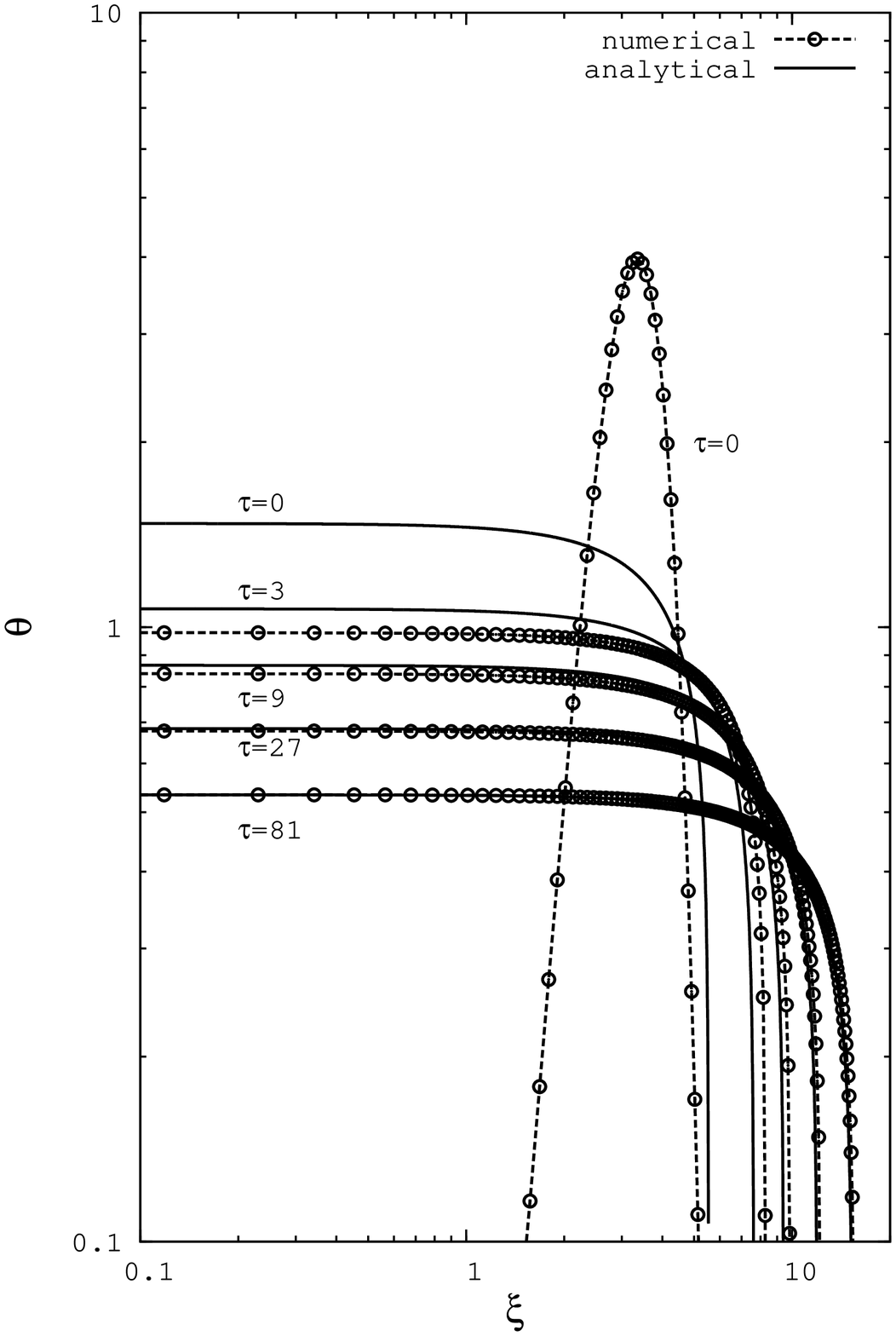}
\includegraphics[width=75mm]{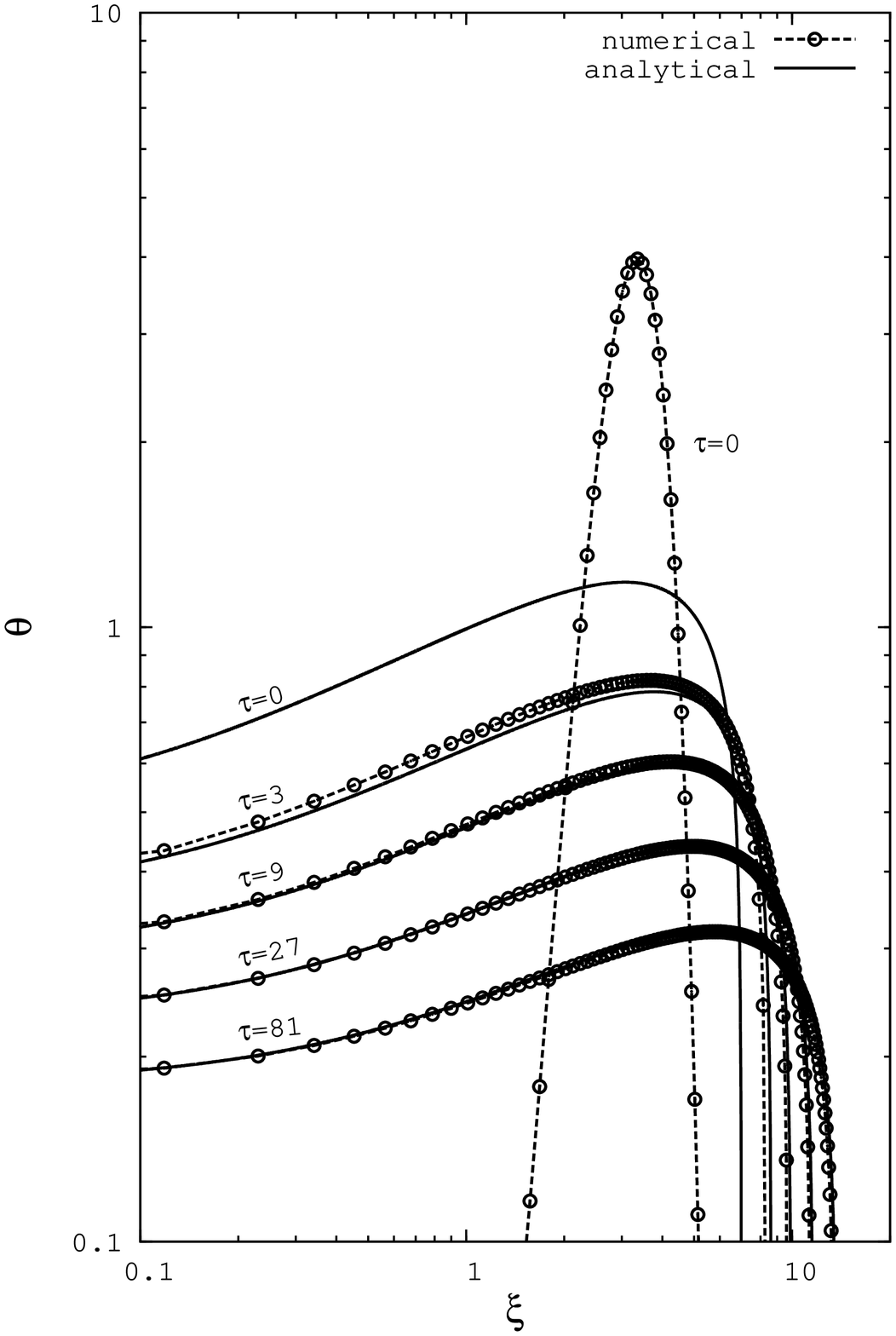}
\caption{Evolution of a Gaussian heat distribution on a semi infinite bar 
described by the diffusion equation (\ref{heat1}), with snapshots taken 
at $\tau=0$, $\tau=3$, $\tau=9$, $\tau=27$ and $\tau=81$ where $\tau$ is 
time in units of diffusive time scale. Solid lines show the analytical solutions 
and dashed lines with data points (corresponding to the numerical grids) show 
the numerical solutions. Left panel corresponds to the case where we employed 
the Neumann boundary condition $\theta'(0,\tau)=0$ and is compared with the 
exact solution given in equation (\ref{sol2a}) with $\tau \rightarrow \tau+1$. 
Right panel corresponds to the case $\theta(0,\tau)=[\Gamma(\tau)]^{1/(n+1)}$ 
and is compared with the
approximate solution given in equation (\ref{unified}).}
\label{fig_unified} 
\end{figure}

The expression
given in equation (\ref{unified}) cannot satisfy the diffusion equation (\ref{heat1}) 
for a general value of $n$ but is always a
good approximate solution. 
In Figure~\ref{fig_unified} we  
compare the self-similar solutions with the numerical solution of the diffusion equation (\ref{heat1}) for 
two different BCs. In both cases we have taken $n=7/3$ and initiated 
the numerical solution from a Gaussian temperature distribution. The analytical
solutions are time shifted as $\tau \rightarrow \tau+1$ exploiting the symmetry of the
diffusion equation (\ref{heat1}) under this transformation. This allows us to evaluate the analytical expression 
at $t=0$. The left panel shows the case with $\theta'(0,\tau)=0$ where the 
exact solution given in equation (\ref{sol2a}) is compared with the numerical solution. 
It is seen that it takes about $\sim 10$ diffusive time scales for the numerical solution 
to forget its initial configuration and to settle onto the self-similar solution. 
The right panel shows a similar comparison for the approximate analytical solution
given in equation (\ref{unified}) for $\Gamma_0=0.1$, $\Phi_0=1.0$. We see that the approximate solution
is also remarkably accurate after $\sim 10$ diffusive time scales. The analytical approximate solution
is even more successful if $n<1$.

\section{DISCUSSION}

We have constructed an approximate solution of the nonlinear heat equation with 
power-law nonlinearity  satisfying a more general BC, by using a ``nonlinear superposition'' of two 
self-similar solutions endowed with Dirichlet and Neumann BCs.
We can not expect to find an exact solution
for such a general BC introduces a length scale into the
problem, thus, destroying the very condition for self-similarity. 
Requiring equation (\ref{unified}) be a solution of equation (\ref{heat1}) reduces to the requirement that
\begin{equation}
 \left( \frac{\tau \dot{\Phi}}{\Phi }+\alpha \right) \xi ^{2}\Phi
^{2}+\left( \frac{\tau \dot{\Gamma}}{\Gamma }+\frac{\tau \dot{\Phi}}{\Phi }%
+2\right) \xi \Gamma \Phi +\left( \frac{\tau \dot{\Gamma}}{\Gamma }+\beta
\right) \Gamma ^{2} \label{require}
\end{equation}
vanishes. The first and the third terms
vanish by equations (\ref{phi1a}) and (\ref{gamma1a}), respectively. The second
term, noting that $2-\alpha-\beta=k_n$, 
does not vanish but simplify to $k_n\xi \Gamma \Phi $. 
This vanishes for either 
$\Gamma_0=0$ or $\Phi_0=0$ corresponding to the well known self-similar solutions
given in Eqns.(\ref{sol1a}) and (\ref{sol2a}). It also vanishes for the linear case, $n=0$, as $k_0=0$. 
The expression given in equation(\ref{unified}) does not satisfy 
the diffusion equation (\ref{heat1}) in general, but it
is a very accurate approximate self-similar solution of it.

The procedure can be applied to other 
second order nonlinear PDEs as long as they have two self-similar solutions and the equation 
can be written in terms of two PDEs in conservative form. 
The application of this procedure to the viscous evolution of a thin accretion disk in the gravitational field 
of a central star, a ubiquitous phenomena in astrophysics \cite{FKR}, will be published elsewhere later. 

Equation (\ref{require}) vanishes when the nonlinearity parameter, $n$, vanishes 
(note that $k_0=0$). As the ``nonlinear superposition'' technique we employed here reduces to ordinary summation
for $n=0$, this gives a way of seeing why summing up two solutions in the linear case yields a solution even when a scale is present: the nonlinearity term $n$ couples with the finite scale in the problem and vanishing of either yields a solution.
It is interesting to see that superposing two solutions by addition is a special case 
of a more general procedure employed in reaching equation(\ref{unified}) from the solutions given by equations (\ref{sol1a}) and (\ref{sol2a}) although this general procedure does not necessarily give an exact solution.

\begin{acknowledgments}
KYE acknowledges Ay\c{s}e Erzan, Sava\c{s} Arapo\u{g}lu, \"{O}mer \.{I}lday, Ahmet T. Giz and \"{O}mer G\"{u}rdo\u{g}an 
for useful discussions and suggestions.
\end{acknowledgments}


\begin{thebibliography}{99}


\bibitem[]{bar96} Barenblatt, G.I. \emph{Scaling, Self-Similarity, and Intermadiate
  Asymptotics} (Cambridge Univ.\ Press, Cambridge, 1996).

\bibitem[]{ray15a} Lord Rayleigh, The principle of similitude. \emph{Nature} {\bf 95}, 66--68 (1915).

\bibitem[]{buck15} Buckingham, E. The principle of similitude. Nature {\bf 96}, 396–-397 (1915).

\bibitem[]{bar72} Barenblatt, G.I. \& Zel'dovich, Y. Self-Similar Solutions as Intermediate Asymptotics.
  \emph{Annu.\ Rev.\ Fluid Mech.} {\bf 4}, 285--312 (1972).

\bibitem[]{dudley} Dudley, J.M., Finot, C., Richardson, D.J. and Millot, G. 
Self-similarity in ultrafast nonlinear optics. \emph{Nature}, {\bf 3}, 597-603 (2007).

\bibitem[]{zel67} Zel'dovich Ya. B. \& Raizer, Yu.P.
  \emph{Physics of Shock Waves and High Temperature Phenomena}
  (Dover Publications, 2002).

\bibitem[]{vasq} V\'{a}zquez, J.L. \emph{Porous Medium Equation} (Oxford Science Publications, 2006).

\bibitem[]{bar52} Barenblatt, G.I. On some unsteady motions of a liquid or a gas in
a porous medium. \emph{Prikl.\ Mat.\ i Mekh.}, {\bf 16} 67-78 (1952).
  

\bibitem[]{bar57}
Barenblatt, G.I. \& Zel'dovich, Ya.B. On the dipole type solution in problems of unsteady gas filtration in the polytropic regime \emph{Prikl.\ Mat.\ i Mekh.}, {\bf 21}, 718-720, 1957.

\bibitem[]{zel59} Zel'dovich Ya.B. \& Kompaneets A.S. On the propagation of heat for nonlinear heat conduction, in
  \emph{Collection dedicated to the seventieth Birthday of
  Academician A.F. Ioffe} (P.I. Lukirskii, ed.) Izdat. Acad. Nauk SSSR, Moskow (1959).

\bibitem[]{FKR} Frank, J., King A. and Raine, D.
  \emph{Accretion Power in Astrophysics} (Cambridge University Press, 2002).

\end{thebibliography}
\end{document}